\documentstyle[aps,prb]{revtex} 
\begin{document}

\twocolumn[\hsize\textwidth\columnwidth\hsize\csname
@twocolumnfalse\endcsname

\title{Electronic response of aligned multishell carbon nanotubes}
\author{J. M. Pitarke$^{1,2}$ and F. J. Garc\'\i a-Vidal$^3$}
\address{$^1$ Materia Kondentsatuaren Fisika Saila, Zientzi Fakultatea, 
Euskal Herriko Unibertsitatea,\\ 644 Posta kutxatila, 48080 Bilbo, Basque
Country, Spain\\
$^2$ Donostia International Physics Center (DIPC) and Centro Mixto
CSIC-UPV/EHU,\\ Donostia, Basque Country, Spain\\
$^3$ Departamento de F\'\i sica Te\'orica de la Materia
Condensada, Facultad
de Ciencias,\\ Universidad Aut\'onoma de Madrid,
28049 Madrid, Spain}

\date\today

\maketitle

\begin{abstract}
We report calculations of the effective electronic response of aligned
multishell carbon nanotubes. A local graphite-like dielectric tensor is
assigned to every point of the multishell tubules, and the effective
transverse
dielectric function of the composite is computed by solving Maxwell's
equations.
Calculations of both real and imaginary parts of the effective dielectric
function
are presented, for various values of the filling fraction and the ratio of
the internal
and external radii of hollow tubules. Our full calculations indicate that
the experimentally measured macroscopic dielectric function of carbon
nanotube
materials is the result of a strong electromagnetic coupling between the
tubes,
which cannot be accounted for with the use of simplified effective medium
theories.
The presence of surface plasmons is investigated, and both optical
absorption cross
sections and energy-loss spectra of aligned tubules are calculated.
\end{abstract}
]

\narrowtext

The so-called multishell carbon nanotubes, with diameters on the
order of
$10\,{\rm nm}$ and lengths reaching up to several microns, are needle-like
tubes
consisting of a finite number of concentric cylindrical two-dimensional
layers of
graphite sheet arranged around a central hollow with a constant separation
between the
layers.\cite{Ijima1,Ebbensen1} Since their discovery,\cite{Ijima1} these
tubules
have attracted much attention because of their size and their novel
structural and
electronic properties.\cite{Ijima2,Dresselhaus,Mintmire,Hamada,Ge} An
important
step in the investigation of these new materials was achieved when
composites were
manufactured with nanotubes organized into well-aligned
arrays,\cite{Ajayan,Heer1}
since alignment of nanotubes was expected to be important not only to test
properties
but also for potential applications such as atomic-scale field
emitters. With the availability of aligned carbon nanotube films, optical
measurements were carried out with polarized light,\cite{Heer1,Heer3}
thereby evaluating the frequency-dependent effective dielectric function of
the composite and showing that carbon nanotubes have an intrinsic and
anisotropic metallic behaviour. Accurate calculations of the effective
dielectric function and the optical response of densely packed carbon nanotubes,
as obtained by solving Maxwell's equations
with the use of tensor-like dielectric functions, have been carried out
only very recently.\cite{fj1}

In this paper, we extend the calculations reported in Ref.\onlinecite{fj1} to
the case of hollow-core nanotubes, and emphasize the effects of the inner
radius on the effective dielectric function of the nanotube array. Calculations
of the so-called energy-loss function, i.e., the imaginary part of the effective
inverse dielectric function, are also presented, for various values of the
filling fraction and the ratio of the internal and external radii of the tubules.

We take a periodic array of infinitely long multishell nanotubes of inner and
outer radii
$r$ and $R$, respectively, arranged in a square array with lattice constant
$a=2\,x\,R$, as shown in Fig. 1. These tubules are assumed to be embedded
in vacuum. In the energy range $0$ to $10\,{\rm eV}$ of
interest in the interpretation of absorption cross sections
and energy-loss spectra, the diameter of typical multishell carbon
nanotubes ($2\,R\sim
10\,{\rm nm}$) is small in comparison to the wavelength of light. Hence,
our results
will be independent of the actual value of the outer radius of the
cylinders. We also
assume that this radius is large enough that a macroscopic dielectric
function is
ascribable to the coaxial cylinders. These cylinders occupy a fraction
$f=\pi(1-\rho^2)/(4\,x^2)$ of space, with $\rho=r/R$ being the ratio
between the inner
and outer radii of the tubes. For simplicity, the magnetic permeabilities
will be
assumed to be equal to unity in all media.

Planar graphite is a highly anisotropic material, and the dielectric
function is a
tensor. This tensor may be diagonalized, by choosing cartesian coordinates
with two
of the axis lying in the basal plane and the third axis being the so-called
$c$-axis. One defines the dielectric function $\varepsilon_\perp(\omega)$
perpendicular to the $c$-axis and the dielectric function
$\varepsilon_\parallel(\omega)$ for the electric field parallel to the
$c$-axis. For
carbon nanotubes, we assume full transferability of the dielectric tensor
of planar
graphite to the curved geometry of carbon tubules, as suggested by Lucas
{\it et
al\,}\cite{Lucas} for the case of multishell fullerenes. Hence, we simply
assign a
local graphite-like dielectric tensor to every point inside the nanotube
and outside
the inner core,
\begin{equation}\label{eq1}
\hat{\varepsilon} (\omega)=\varepsilon_{\perp}(\omega)
({\bf \theta}{\bf \theta}+{\bf z}{\bf z})+
\varepsilon_{\parallel}(\omega){\bf r}{\bf r},
\end{equation}
${\bf \theta}{\bf \theta}$, ${\bf z}{\bf z}$, and ${\bf r}{\bf r}$
being the unitary basis vectors of cylindrical coordinates, and take the
principal dielectric functions $\varepsilon_\perp(\omega)$ and
$\varepsilon_\parallel(\omega)$ of graphite from Ref.\onlinecite{Palik}.

In the long-wavelength limit, a composite material may be treated as if it
were
homogeneous, with the use of an effective dielectric function
$\varepsilon_{eff}$.
The optical absorption cross section of the composite is then directly
given by
${\rm Im}\,\varepsilon_{eff}(\omega)$. Also, for small values of the
adimensional
parameter $qR$ ($qR<1$), $q$ being the momentum transfer, the energy-loss
spectra of a broad beam of swift electrons penetrating the composite is
found\cite{Pitarke1} to be well described by the $q\to 0$ limit of
the imaginary part of the effective longitudinal dielectric
function,\cite{note1} i.e., the
so-called energy-loss function, ${\rm
Im}[-\varepsilon_{eff}^{-1}(\omega)]$.  

In order to compute, with full inclusion of the electromagnetic interaction
between the tubules, the effective dielectric function of our periodic system, we
first solve Maxwell's
equations. The local dielectric function of our composite material depends
on
frequency. Thus, on-shell methods for solving Maxwell's
equations,\cite{Pendry1} which
proceed at constant frequency, are particularly well suited for our
calculations.
Though the methodology of Ref.\onlinecite{Pendry1} was originally reported
to solve
Maxwell's equations with scalar dielectric functions, an extension of this
formalism
was developed in Ref.\onlinecite{Pendry2}, which enables us to consider
tensor-like
dielectric functions like the one of Eq. (\ref{eq1}).

We first fix the frequency, thereby the local tensor-like dielectric function at
any point of the composite material being specified, and approximate the
continuous electromagnetic field by its values at a series of mesh points located
on a simple square lattice with unit cell of side $a$, as shown in Fig. 1. The
corresponding discretized Maxwell's equations provide a relationship between the
electromagnetic fields on either side of the unit cell, i.e., $x$ and $x+a$ (see
Fig. 1). We exploit this to calculate, on applying Bloch's
theorem, the eigenvalues of the so-called transfer matrix, which will give
the band structure of the system, i.e., the dispersion relations $k(\omega)$ of
the Bloch-state wave vector of propagation versus the frequency. The effective
transverse dielectric function is then obtained as follows,
\begin{equation}\label{eq3}
\varepsilon_{eff}(\omega)={k^2(\omega)c^2\over\omega^2},
\end{equation}
where $c$ represents the speed of light.

Fo an e.m. wave normally incident on the structure ($k_y=k_z=0$) there are two
different values of $\varepsilon_{eff}(\omega)$ corresponding to $s$ and $p$
polarizations. For e.m. waves polarized along the tubes ($s$ polarization), the
electric field is parallel to the cylinders at every point, and is not
modified
by the presence of the interfaces. Hence, the effective dielectric function
of the
composite is simply the weighted average of the dielectric functions of the
constituents,\cite{Bergman}
\begin{equation}\label{eq4}
\varepsilon_{eff}(\omega)=f\,\varepsilon_\perp(\omega)+(1-f).
\end{equation}
Our numerical results for the effective dielectric function, as obtained from Eq.
(\ref{eq3}) for this polarization, accurately reproduce
the exact results predicted by Eq. (\ref{eq4}), which represents a good
check for our
scheme. According to Eq. (\ref{eq4}), ${\rm
Im}\,\varepsilon_{eff}(\omega)=f\,{\rm
Im}\,\varepsilon_\perp(\omega)$; therefore, the shape of the experimentally
determined
${\rm Im}\,\varepsilon_{eff}(\omega)$ must coincide for this polarization
with ${\rm
Im}\,\varepsilon_\perp(\omega)$. The measurements of
$\varepsilon_{eff}(\omega)$
reported in Ref.\onlinecite{Heer1} for light polarized along the tubes are
plotted in
Fig. 2 by solid circles, together with the effective dielectric function
predicted by Eq. (\ref{eq4}),
which roughly reproduces with $f\sim 0.5$ the experimentally determined
${\rm
Im}\,\varepsilon_{eff}(\omega)$. We note, however, that neither the exact
position of
the maximum nor the details of the real part of the measured effective
dielectric
function can be reproduced with the use of Eq. (\ref{eq4}). This may
reflect existing
differences between carbon structures in planar graphite and in nanotubes.
Also, optical measurements may be affected by the
presence of tubes with different orientations.

The effective energy-loss function, ${\rm Im}[-\varepsilon_{eff}^{-1}(\omega)]$,
corresponding
to $s$ polarization is also plotted in Fig. 2. For this polarization, the
effective inverse
dielectric function has a single pole, which occurs at a reduced plasma
frequency
where ${\rm Re}\,\varepsilon_\perp\sim-(1-f)/f$ and ${\rm
Im}\,\varepsilon_\perp\sim
0$. For graphite, at low concentrations of tubules ($f<0.2$) the resonance
condition is never satisfied, and in the limit as $f\to 0$ the effective
energy-loss
function simply coincides with ${\rm Im}\,\varepsilon_\perp(\omega)$, thereby
showing the
characteristic peak at $\sim 4.6\,{\rm eV}$. This peak is associated with the
maximum in the joint density of states of the $\pi$ valence and conduction
bands in graphite.

Calculations of the $p$ component of the effective dielectric function of an
array of plain ($\rho=0$) carbon nanotubes were reported in
Ref.\onlinecite{fj1} for various values of the ratio $x$ between the lattice
constant and the outer diameter of the cylinders, showing that the trend with
increasing the concentration of tubules is for the actual dipolar peak in
${\rm Im}\,\varepsilon_{eff}(\omega)$ to be shifted from the isolated-cylinder
dipole mode at $\sim 6.5\,{\rm eV}$ to lower energies. When the nanoparticles are
brought into close contact, electromagnetic coupling between them converts the
dipolar surface mode into a very localized one, trapped in the region between the
nanostructures. At higher concentrations of tubules, when graphite forms a
connected medium ($x\le 1$), dipolar modes cannot be excited, and
the optical absorption exhibits a single peak originated in the maximum of
${\rm Im}\,\varepsilon(\omega)$ at $\sim 4.6\,{\rm eV}$. We have also
calculated the energy-loss function of an array of plain carbon
nanotubes,\cite{new} and have found that for $x\le 1$ it shows a single peak at
the bulk plasmon resonance at $\sim 7\,{\rm eV}$, where ${\rm
Re}\,\varepsilon_\perp(\omega)$ and ${\rm Im}\,\varepsilon_\perp(\omega)$ are
both small.

For hollow tubes ($\rho\neq 0$), there are two distinct dipolar modes with
either tangential or radial symmetry, similar to those present in the
case of a thin planar film\cite{Ritchie} and a spherical shell.\cite{Lucas0}
With the presence of anisotropy this two-mode structure is replaced by a
more complicated spectral representation.\cite{new} Our full calculations of the
$p$ effective dielectric function of a periodic array of hollow carbon nanotubes,
as obtained for various ratios between inner and outer radii, are shown
in Figs. 3 and 4 with $x=2.0$ and $x=1.03$, respectively.\cite{note3} As in the
case of plain cylinders ($\rho=0$), our full calculations nearly coincide for
$x=2.0$ with a generalized Maxwell-Garnett (MG) effective dielectric function
appropriate for anisotropic hollow tubes.\cite{MG} However, in the close-packed
regime ($x=1.03$) the strong electromagnetic coupling between the tubes yields
non-negligible contributions from multipolar resonances. This multipolar
coupling provokes a redshift and a blueshift of the MG dipolar resonances that
are visible in the optical spectra and the energy loss, respectively. These are
the low-energy dipolar mode with tangential symmetry, which absorbs light (${\rm
Im}\,\varepsilon_{eff}(\omega)$ is maximum at the corresponding energy), and the
high-energy dipolar mode with radial symmetry, which can be excited by moving
charged particles (${\rm Im}[-\varepsilon_{eff}^{-1}(\omega)]$ is maximum at the
corresponding energy).

The experimentally determined macroscopic dielectric function of
close-packed carbon nanotubes, as reported in Ref.\onlinecite{Heer1} for
$p$-polarized light, is exhibited in Fig. 4 by solid circles. A comparison
between these measurements and the $p$
effective dielectric function of closed-packed ($x=1.03$) plain ($\rho=0$)
carbon
nanotubes was presented in Ref.\onlinecite{fj1}, showing a nice agreement
between
theory and experiment, though a shift of approximately 1 was needed to
obtain agreement
with the real part of $\varepsilon_{eff}(\omega)$.\cite{note2} If one chooses the
ratio
$\rho$ between
the inner and outer radii of the tubes to be 0.6, which yields in the
close-packed
regime ($x=1.03$)  a filling fraction $f\sim 0.5$, both real and imaginary
parts of
our calculated effective dielectric function show an excellent agreement
with the
experiment. This filling fraction ($f\sim 0.5$) is precisely what is
required to also
reproduce the dielectric function experimentally determined with the use of
light
polarized along the cylinders, as discussed above (see Fig. 2).

In conclusion, we have shown that the effective dielectric function of
densely packed carbon nanotubes is the result of a
strong electromagnetic coupling between the tubes. We have chosen the inner and
outer diameters of hollow tubules to be $6$ and
$10\,{\rm nm}$, respectively ($\rho=0.6$), and have obtained both $s$ and $p$
effective dielectric functions to be in the close-packed regime ($a=10.3\,{\rm
nm}$, $x=1.03$) in excellent agreement with the experimental measurements. Small
discrepancies have been observed in the case of light polarized along the tubes
($s$ polarization), which may reflect existing differences between the
dielectric function of carbon structures in planar and curved graphite.

J.M.P. gratefully acknowledges partial support by the University of the
Basque Country, the Basque Hezkuntza, Unibertsitate
eta Ikerketa Saila, and the Spanish Ministerio de Educaci\'on y Cultura.

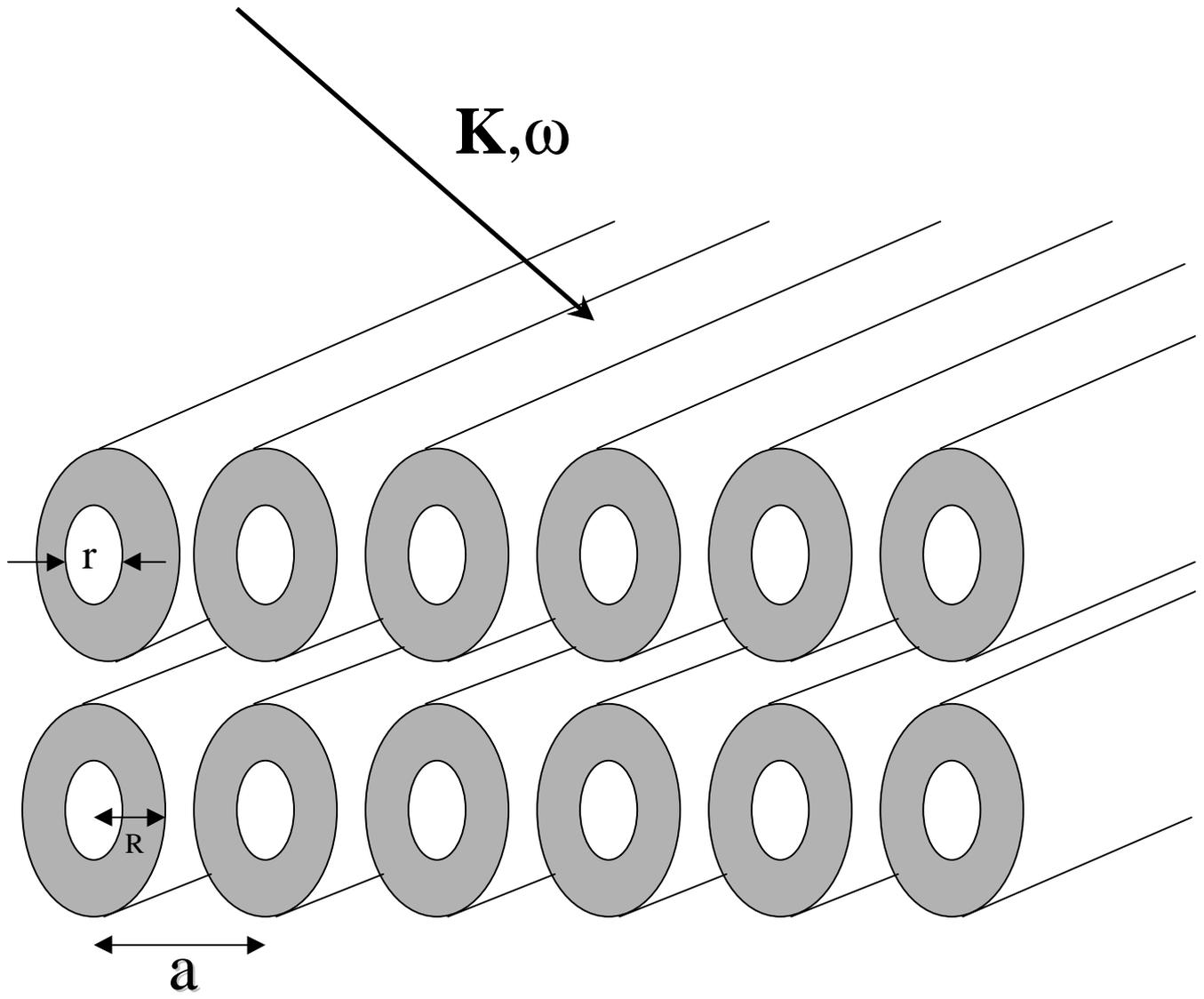
\begin{figure}
\caption{Multishell nanotubes of inner and outer
radii $r$ and $R$, respectively, arranged in a square array with lattice
constant
$a$. The cylinders are infinitely long in the $y$ direction. The e.m.
interaction of
this structure with a normally incident plane wave of momentum ${\bf k}$
[$k_y=k_z=0$] and energy
$\omega$ is investigated.}
\end{figure}

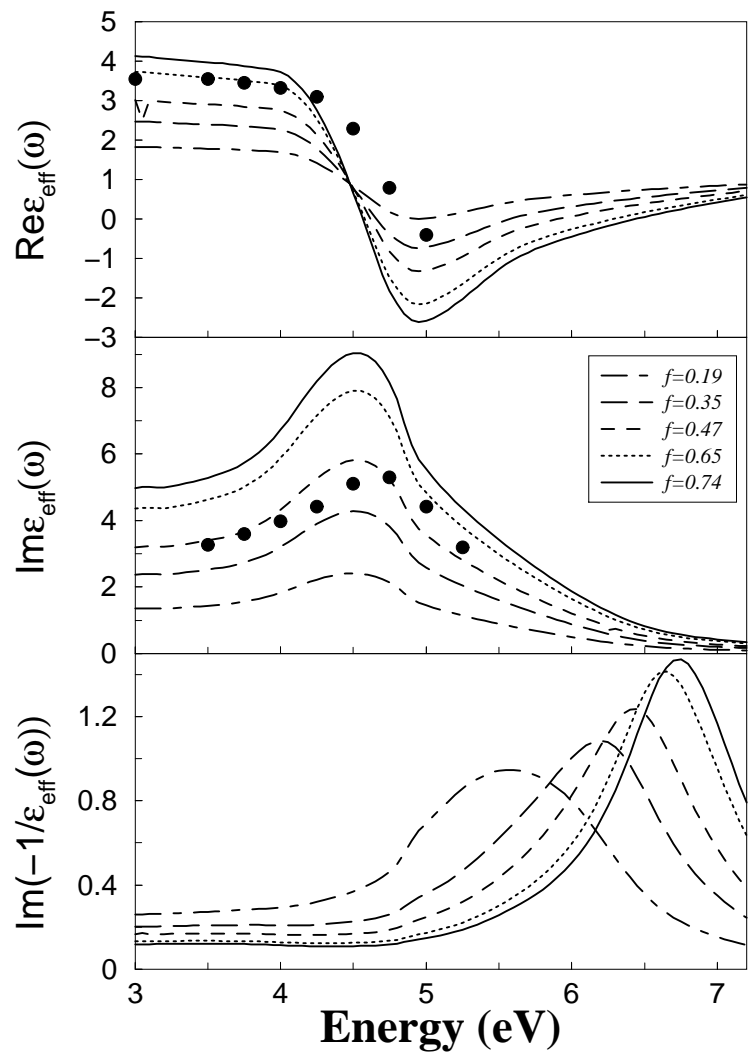
\begin{figure}
\caption{The real and imaginary parts of the long-wavelength effective
dielectric
function, ${\rm Re}\,\varepsilon_{eff}(\omega)$ and ${\rm
Im}\,\varepsilon_{eff}(\omega)$, and the energy-loss function, ${\rm
Im}[-\varepsilon_{eff}^{-1}(\omega)]$, of the periodic system described in
Fig. 1, for
$s$ polarized electromagnetic excitations. Dashed-dotted,
long-dashed, short-dashed, dotted, and solid lines: calculated results, from
either Eq. (\ref{eq3}) or Eq. (\ref{eq4}), for
volume filling fractions of $19\%$, $35\%$, $47\%$,
$65\%$, and $74\%$, respectively. The solid circles represent the
experimental
results reported in Ref.\protect\onlinecite{Heer1}\protect.}
\end{figure}

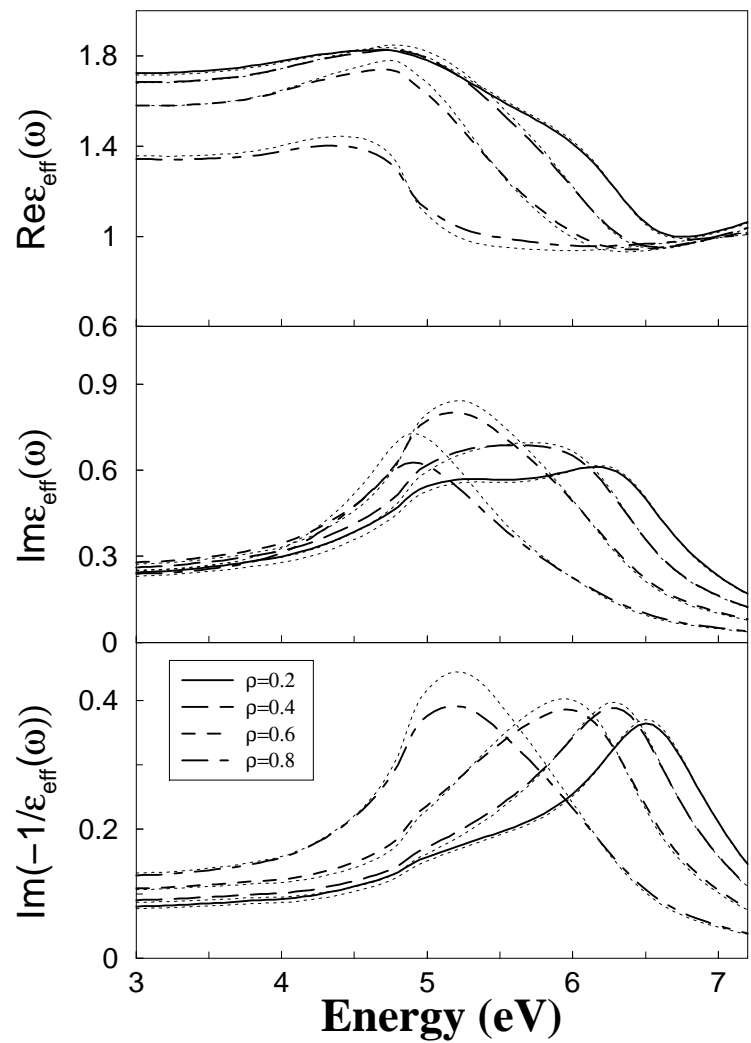
\begin{figure}
\caption{The real and imaginary parts of the long-wavelength effective
dielectric
function, ${\rm Re}\,\varepsilon_{eff}(\omega)$ and ${\rm
Im}\,\varepsilon_{eff}(\omega)$, and the energy-loss function, ${\rm
Im}[-\varepsilon_{eff}^{-1}(\omega)]$, of a periodic array of hollow carbon
nanotubes,
for $p$ polarized electromagnetic excitations and for the ratio between the
lattice
constant and the outer diameter of the cylinders $x=2.0$. Solid, long-dashed,
short-dashed, and dashed-dotted lines represent our full calculations, as
obtained from Eq.
(\ref{eq3}), for ratios between the inner and outer radii of the tubes
$\rho=0.2, 0.4, 0.6$, and $0.8$, respectively. The dotted lines
represent a generalized MG effective dielectric function appropriate for
anisotropic hollow cylinders.}
\end{figure}

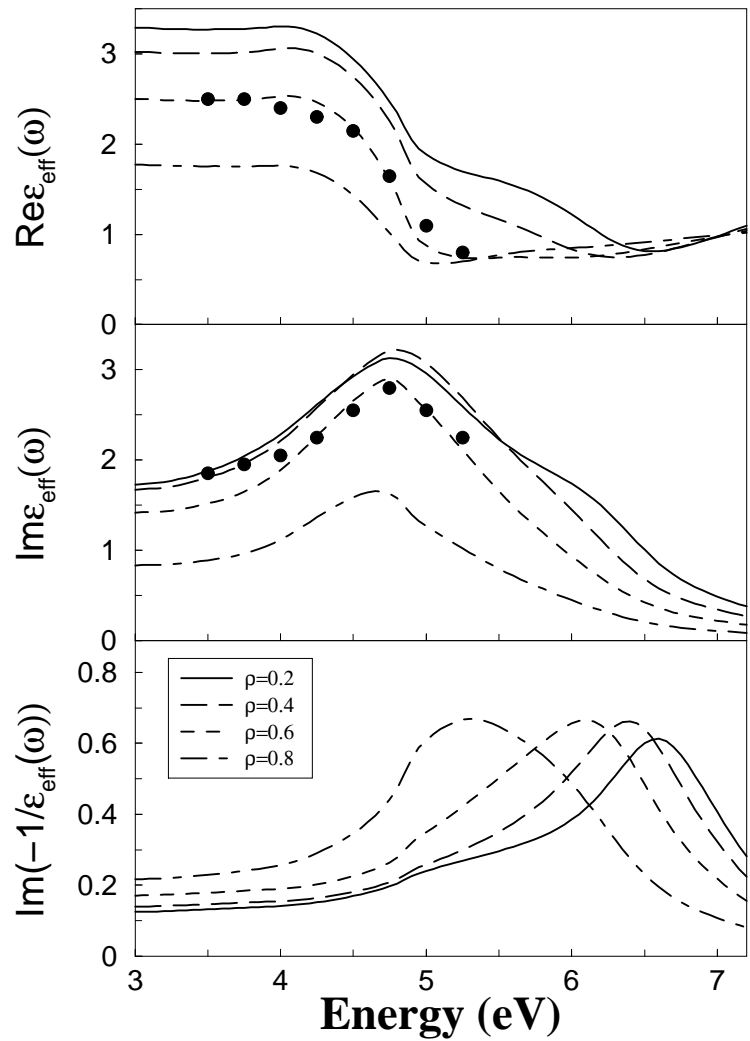
\begin{figure}
\caption{Same as Fig. 3, for the ratio between the lattice constant and
the
outer diameter of the cylinders $x=1.03$. The solid circles represent the
experimental results reported in Ref.\protect\onlinecite{Heer1}\protect.}
\end{figure}


\begin{references}

\bibitem{Ijima1} S. Ijima, Nature {\bf 354}, 56 (1991).
\bibitem{Ebbensen1} T. W. Ebbesen and P. M. Ajayan, Nature {\bf 358}, 220
(1992).
\bibitem{Ijima2} S. Ijima {\it et al}, Nature {\bf 356}, 776
(1992).
\bibitem{Dresselhaus} M. S. Dresselhaus, Nature {\bf 358}, 195 (1992).
\bibitem{Mintmire} J. W. Mintmire {\it et al}, Phys. Rev. Lett.{\bf 68}, 631
(1992).
\bibitem{Hamada} N. Hamada {\it et al}, Phys. Rev. Lett.
{\bf 68},
1579 (1992).
\bibitem{Ge} M. Ge and K. Sattler, Science {\bf 260}, 515 (1993).
\bibitem{Ajayan} P. M. Ajayan {\it et al}, Science {\bf 265}, 1212 (1994).
\bibitem{Heer1} W. A. de Heer {\it et al}, Science {\bf 268}, 845 (1995).
\bibitem{Heer3} F. Bommeli {\it et al}, Solid. State. Commun. {\bf 99}, 513
(1996).
\bibitem{fj1} F. J. Garc\'\i a-Vidal {\it et al},
Phys. Rev. Lett. {\bf 78}, 4289 (1997).
\bibitem{Lucas} A. A. Lucas {\it et al}, Phys. Rev. {\bf 49}, 2888
(1994).
\bibitem{Palik} E. D. Palik, Ed., {\it Handbook of Optical Constants of
Solids} (Academic, New York, 1985).
\bibitem{Pitarke1} J. M. Pitarke {\it et al}, Phys. Rev. B
{\bf 55}, 9550 (1997); J. M. Pitarke and A. Rivacoba, Surf. Sci. {\bf 377},
294 (1997).
\bibitem{note1} In the long-wavelength limit ($q\to 0$), longitudinal and
transverse
dielectric functions with the same polarization coincide.
\bibitem{Pendry1} J. B. Pendry and A. MacKinnon, Phys. Rev. Lett. {\bf 69},
2772
(1992); J. B. Pendry, J. Mod. Opt. {\bf 41}, 2417 (1994).
\bibitem{Pendry2} A. J. Ward and J. B. Pendry, J. Mod. Opt. {\bf 43}, 773
(1996).
\bibitem{Bergman} D. J. Bergman and D. Stroud, Solid State Phys. {\bf 46},
147 (1992).
\bibitem{new} F. Garc\'\i a-Vidal and J. M. Pitarke (unpublished).
\bibitem{Ritchie} R. H. Ritchie, Phys. Rev. {\bf 106}, 874 (1957).
\bibitem{Lucas0} A. A. Lucas {\it et al}, Nucl. Instrum.
Methods B {\bf 96}, 470 (1995).
\bibitem{note3} For metallic structures, sampling meshes
as large as $180\times 180$ have been found to be required to
provide well-converged results;\cite{Pitarke2} however, for carbon nanotubes
sampling meshes of $60\times 60$ have been found to provide well-converged
results, which is  due to the smoothing effect of the large damping originated
with the presence of interband transitions in graphite.
\bibitem{Pitarke2} J. M. Pitarke {\it et al}, Phys. Rev. B {\bf 57}, 15261
(1998); J. M. Pitarke {\it et al}, Surf. Sci. {\bf 433}, 605 (1999). 
\bibitem{MG} J. C. Maxwell-Garnett, Philos. Trans. R. Soc. London A {\bf
203}, 385 (1904); {\bf 205}, 237 (1906); see, also, C. F. Bohren and D. R.
Huffman, {\it Absorption and Scattering of light by Small Particles} (Wiley, New
York, 1983).
\bibitem{note2} In Fig. 3 of Ref.\onlinecite{fj1} the curve labeled (a) for the
real part of the effective dielectric function was shifted by a constant value
of $+4$, while the experimental data for this quantity were shifted by a
constant value of $+5$.
\end{references}
\end{document}